# Renewable energy exporting consumption-oriented transfer limit switching control: A unsupervised learning-based method

Gao Qiu, Haojin Peng, Youbo Liu, Tingjian Liu, Junyong Liu
College of Electrical Engineering
Sichuan University
Chengdu, Sichuan, China
qiugaoscu@stu.scu.edu.cn, penghaojin@stu.scu.edu.cn, liuyoubo@scu.edu.cn, scueeliutj@163.com, liujy@scu.edu.cn

*Abstract*—A method for generating unsupervised conditional mapping rules for multi-inter-corridor transfer limits and their integration into unit commitment through banding-switching is proposed in this paper. The method starts by using Ant colony clustering(ACC) to identify different operating modes with renewable energy penetration. For each sub-pattern, coupling inter-corridors are determined using correlation coefficients. An algorithm for constructing coupled inter-corridors' limits boundaries, employing grid partitioning, is proposed to establish conditional mappings from sub-patterns to multi-inter-corridor limits. Additionally, a banding matching model is proposed, incorporating distance criteria and the Big-M method. It also includes a limit-switching method based on Lagrange multipliers. Case studies on the IEEE 39-node system illustrate the effectiveness of this method in increasing consumption of renewable energy and reducing operational costs while adhering to stability verification requirements.

*Index Terms*-- Multiple inter-corridors' coupling power transfer limits, conditional mapping rule, unsupervised learning

## I. INTRODUCTION

To ensure power grid safety and stability without adding unnecessary complexity, grid operators often use the most conservative transfer limits from multiple sets of calculated total transfer capabilities (TTC) under various extreme scenarios[1], [2]. These conservative limits are subsequently utilized to set constraints on power transmission within inter-corridors. It's crucial to understand that, even though TTC calibration can be applied to establish these inter-corridor limits, the two concepts, TTC and inter-corridor limits, may appear similar but are fundamentally distinct.

TTC represents the maximum transfer capability of individual inter-corridor within the stable operation mode set defined by inter-corridor cuts. It's a dynamic parameter that varies with operating conditions. A single TTC cannot serve as a fixed stability boundary for inter-corridor. They are often impractical for two main reasons. Firstly, common methods, including continuous power flow[3], repeated power flow[4], and optimal power flow[5], struggle to provide real-time solutions for calculating TTC, especially when dealing with various stability constraints, N-k constraints, voltage-reactive constraints, and other safety constraints. This makes them unsuitable for fast stability assessments in highly dynamic renewable energy grids. Even AI-based solutions face challenges in terms of engineering applicability and model reliability[6], [7]. Secondly, even if there were methods to address TTC calculation efficiency issues, the resulting dynamic safety boundaries are hard to apply to other static power market computations and are not compatible with existing mature solvers[8].

On the other hand, inter-corridor limits represent the minimum power level within the instability mode set, defining the most conservative safety envelope under inter-corridors' power transfer cutsets. These limits are fixed parameters, addressing the computational complexity issues encountered with TTC. While the online application of inter-corridor limits is highly efficient, their calibration process faces some critical challenges. Firstly, similar to TTC, inter-corridor limit calculation remains a high-dimensional, non-convex, nonlinear problem combining multiple safety and stability constraints. Secondly, there is currently no comprehensive research bridging the scientific calculation of multi-inter-corridor limits with optimization control. These issues persistently confine real-world limit calculations to traditional engineering paradigms, where precise validation remains elusive, and conservatism is often overemphasized.

To address these challenges, this paper introduces a data-driven unsupervised method for generating multi-inter-corridor limits condition mapping rules and embedding them into a unit commitment model for optimization. Firstly, the ACC algorithm is employed to identify a set of typical grid operating modes, emphasizing the boundaries between stable and unstable modes. Subsequently, each sub-operating mode is examined, and a grid-based method is proposed for constructing multi-inter-corridor coupled limits boundaries. This allows for the offline calculation of "sub-operating mode-limit" condition mapping rules. Finally, a limit switching control model is developed based on Euclidean distance and the Big-M method, integrated into the unit commitment optimization framework using the Lagrange multiplier method.

Case studies conducted on the IEEE 39-node system validate the advantages of the proposed method in enhancing the efficiency of renewable energy export in multi-inter-corridor.

## II. UNIT COMMITMENT WITH INTER-CORRIDORS' LIMITS

When considering the combination of generating units subject to inter-corridor transmission limit constraints, our objective is to minimize the generation costs of conventional units and the curtailment costs of renewable energy units (in this paper, the renewable energy units are limited to wind power units). The decision variables encompass power generation levels and the on/off states of the units, as follows:

$$\underset{x}{\text{Minimize }} C(x) \quad (1)$$

$$\begin{cases} \text{s.t. } Z(x, y) = 0 \\ \quad\; F(x, y) \leq 0 \end{cases} \quad (2)$$

$$x = [P_g, u_g, P_w] \quad (3)$$

where $\mathbb{G}, \mathbb{W}$ represent the sets of conventional generating units and wind power units, respectively $(\forall g \in \mathbb{G}, w \in \mathbb{W})$. $C(x)$ represents the quadratic fuel cost function for conventional units. $x, y$ represent the system control variables and state variables, where $x$ includes $P_g, u_g, P_w$, representing the output of conventional units, the on/off status of conventional units, and the output of wind power units, respectively. $Z(\cdot)$ and $F(\cdot)$ represent the equality and inequality constraints within the unit commitment. These constraints encompass generator output limits, power balance constraints, startup and shutdown time constraints, system reserve constraints, as well as generator ramping constraints [9].

In practice, ensuring the security of power grid operations requires the consideration of inter-corridor transmission limit constraints. Unlike dynamic TTC constraints, transmission limits are static values, resulting in constraints with lower complexity. They are more suitable for practical engineering applications and are detailed as follows:

$$P_s(t) = \sum_{l \in \mathbb{L}^s} \left( \sum_{g \in \mathbb{G}} G_{g-l} P_g(t) + \sum_{w \in \mathbb{W}} G_{w-l} P_w(t) - \sum_{d \in \mathbb{D}} G_{d-l} P_d(t) \right) \quad (4)$$

$$\Gamma_s^-(t) \leq P_s(t) \leq \Gamma_s^+(t) \quad (5)$$

where $\mathbb{S}$ represents the set of inter-corridors; $\mathbb{L}^s$ represents the set of transmission lines connected to inter-corridor $s$ ($\forall s \in \mathbb{S}$); $G_{i-l}$ denotes the power transfer factor from the $i$-th generator/wind turbine/load to line $l$; $P_s(t)$ is the flow value of inter-corridor $s$ at time period t; $\Gamma_s^-(t)$ and $\Gamma_s^+(t)$ represent the limit values for the forward and reverse power transmission of inter-corridor $s$ at time period $t$. It's important to note that (5) describes the single-inter-corridor limit constraint without considering the inter-inter-corridor coupling.

## III. UNSUPERVISED CONDITIONAL MAPPING RULE GENERATION FOR MULTIPLE POWER TRANSFER INTER-CORRIDORS' COUPLING LIMITS

### A. Unsupervised Learning-Based Operating Mode Identification

To address the conservatism in traditional limit calculations, this paper proposes an inter-corridor limit condition mapping model based on unsupervised learning (UL). In particular, the paper begins by employing the ACC algorithm for fine-grained identification of operating mode clusters[10], [11]. The selected input features include synchronous generator active power $P_g$, wind generator active power $P_w$, and load active power $P_d$. The on/off state of the generators is implicitly encoded in the output power (output power of 0 indicates shutdown). To simplify the description, the notation $o \in \mathbb{R}^{n \times l}$ is used to represent input features, where $n$ and $l$ denote the sample size and feature dimensions, respectively.

$$o = [o_1; \ldots; o_i; \ldots; o_n], \forall o_i \in \mathbb{R}^{1 \times l}, o \subseteq \mathbb{E} \quad (6)$$

$$o_i = [P_g, P_w, P_d], \forall g \in \mathbb{G}, w \in \mathbb{W}, d \in \mathbb{D} \quad (7)$$

$$\mathcal{C} = [c_1; \ldots; c_j; \ldots; c_k], \forall c_j \in \mathbb{R}^{1 \times l}, c_j \subset o, k = |\mathbb{J}| \quad (8)$$

where $\mathbb{D}$ represents the set of loads. $\mathcal{C}$ represents the matrix of class center operating modes, and $\mathbb{J}$ represents the set of operating mode clusters, with $k$ denoting the number of clusters. Subsequent steps involve tuning the corresponding inter-corridor transmission limits for each cluster, resulting in a conditional mapping set of size $|\mathbb{J}|$ hence $k=|\mathbb{J}|$. For the sake of clarity, this paper simultaneously defines the number of tiers for conditional mapping rules, corresponding to the size of the conditional mapping set. In other words, the number of tiers for conditional mapping rules is equal to the number of clusters, which is $k=|\mathbb{J}|$.

### B. Grid-Based Generation of Multi-Inter-Corridor Coupled Limit Conditional Mapping Rules

In the case of multi-inter-corridor transmission limits with coupling effects, not considering the correlation between inter-corridors may reduce the feasibility of dividing the multi-inter-corridor limits or even lead to the problem of having no feasible solutions. Therefore, it is necessary to describe multi-inter-corridor limits with logical association constraints based on the consideration of correlation. Specifically, the first step involves calculating the correlation between inter-corridors using the Pearson correlation coefficient(PCC)[12] and identifying the inter-corridors with strong correlations.

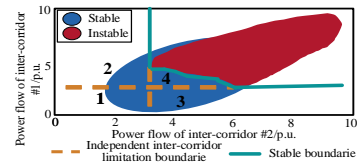

Figure 1. The security boundary for multi-inter-corridor coupled power transfer limits

Figure 1 illustrates a set of positively correlated inter-corridor flow samples. The feasible region obtained through the traditional limit calculation model is represented as Region 1, which discards the potentially useful stable operation regions, such as Region 2 and Region 3. It is evident that one major factor contributing to the conservatism of traditional limits is the simplistic inter-corridor of independent inter-corridor limits in engineering, forming a conservative feasible region. Furthermore, engineering algorithms often neglect the stable mode region, such as Region 4, determined by the joint distribution of inter-corridor information. To address these issues and avoid the generation of overly complex limit constraints while adhering to engineering conservatism, interpretability, and well-defined boundaries, this paper proposes a grid partition-based method for generating multi-inter-corridor transmission limit constraints. The steps are as follows:

① Initialize all critical inter-corridor transmission limit values using (9) and employ the PCC to screen for correlated pairs of inter-corridors.

$$\begin{cases} \overline{\Gamma_{s,j}} = \min\left(\underline{P_{s,is,j}}, \overline{P_{s,se,j}}\right) \\ \underline{P_{s,is,j}} = \min_{\forall i \in \mathbb{E}_j \cap \{i | S_h^{(i)}=0\}} P_s^{(i)}, \forall h \in \mathbb{H}, s \in \mathbb{S}, j \in \mathbb{J} \\ \overline{P_{s,se,j}} = \max_{\forall i \in \mathbb{E}_j \cap \{i | S_h^{(i)}=1\}} P_s^{(i)}, \forall h \in \mathbb{H}, s \in \mathbb{S}, j \in \mathbb{J} \\ \text{s.t.} \quad \mu_{s,se,j} < \mu_{s,is,j} \end{cases}$$ (9a)

$$\begin{cases} \underline{\Gamma_{s,j}} = \max\left(\overline{P_{s,is,j}}, \underline{P_{s,se,j}}\right) \\ \overline{P_{s,is,j}} = \max_{\forall i \in \mathbb{E}_j \cap \{i | S_h^{(i)}=0\}} P_s^{(i)}, \forall h \in \mathbb{H}, s \in \mathbb{S}, j \in \mathbb{J} \\ \underline{P_{s,se,j}} = \min_{\forall i \in \mathbb{E}_j \cap \{i | S_h^{(i)}=1\}} P_s^{(i)}, \forall h \in \mathbb{H}, s \in \mathbb{S}, j \in \mathbb{J} \\ \text{s.t.} \quad \mu_{s,se,j} > \mu_{s,is,j} \end{cases}$$ (9b)

where $\mathbb{H}$ represent the anticipated contingency set, which includes static, transient, and N-k faults. $\mathbb{S}$ represents the set of transmission inter-corridors. $S_h^{(i)}$ indicates the safety label after safety verification for the $i$th scenario and $h$th anticipated contingency, where 0 signifies unsafe, and 1 indicates the opposite. $\mathbb{E}$ denotes the set of operating modes used for limit calculations. $\mu_{s,se,j}$ and $\mu_{s,is,j}$ represent the expected power values of $s$th inter-corridor corresponding to the stable and unstable operating modes in the $j$th operating mode cluster. The overline and underline denote upper and lower bounds, respectively. The initial inter-corridor limit calculated by $\mu_{s,se,j} < \mu_{s,is,j}$ is the upper bound of the stable region, and vice versa, it is the lower bound.

② Here is the algorithm flow for generating multi-inter-corridor limit constraints based on the grid approach, considering a pair of inter-corridor a and b:
1) Input sample power data for inter-corridor a and inter-corridor b and determine the grid resolution.
2) Define a rectangular domain for inter-corridor power based on the boundaries of unstable samples. The domain's boundaries are determined by the upper and lower limits of power for unstable samples.
3) Calculate the search step sizes in the direction of inter-corridor a and inter-corridor b by dividing the length of the rectangular domain sides by their respective resolutions.

4) Perform a breadth-first search on grids corresponding to safe operating conditions only, determining the coupled limit boundaries. Specifically, iterate through grids along the step sizes of inter-corridor a and inter-corridor b, nested within a loop, and update the boundaries. Continue until a grid containing an unstable mode is found, then stop the search and record the boundaries.
5) Finally, compute the union of the obtained grid boundaries with the initial boundaries from step 1 to obtain an approximate limit boundary set, denoted as $\tilde{\partial}\Omega_j$.

The aforementioned steps are deployed in the offline phase. Once the mapping rules for inter-corridor limit conditions are obtained, they are embedded into the scheduling model during the online phase. Based on operational condition information, the unique cluster to which it belongs is determined, serving as a criterion to provide the corresponding inter-corridor limit mapping. Hence, the rules belong to a typical "if-then" relationship data structure[13], and its mathematical model is as follows:

$$j = \arg\min_{j=1,\ldots,k} \| \varphi(\boldsymbol{x}(t)) - \boldsymbol{c}_j \|_2 \quad (10a)$$

$$\boldsymbol{\Gamma}(t) = f(\boldsymbol{x}(t)): \boldsymbol{x}(t) \in \mathbb{E}_j \rightarrow \boldsymbol{\Gamma}(t) = \tilde{\partial}\Omega_j \quad (10b)$$

where $\boldsymbol{\Gamma}(t)$ represents the inter-corridor limit constraint for period $t$, $x(t)$ denotes the operating mode variables for period $t$, and $\varphi(t)$ stands for a feature extractor used to extract features from the current operating mode, which are used to determine the subset of sub-operating modes to which it belongs. Equation (10a) determines the membership set $\mathbb{E}_j$ by solving the minimization problem with the least square norm between $\varphi(\boldsymbol{x}(t))$ and $c_j$, and the mapping from operating modes to multiple inter-corridor limit constraints is achieved through the defined conditional mapping $f(\cdot)$.

IV. TIERED SWITCHING MODEL AND SOLUTION PARADIGM FOR MULTI-INTER-CORRIDOR COUPLED LIMITS

The aforementioned algorithm addresses the issue of conservative settings in traditional inter-corridor limit constraints and forms controllable multiple inter-corridor limit switching rules (10) based on clustering affiliation criteria (i.e., Euclidean distance). However, a method for integrating these rules into the power grid scheduling model is currently lacking. In this paper, we focus on the unit commitment problem under inter-corridor limit constraints and propose an optimization paradigm that guides inter-corridor limit switching based on a two-tier mixed-integer programming approach.

We modify the unit commitment model considering inter-corridor limit constraints as follows: by replacing the fixed limit constraints in (5) with (10), we establish an upper-level model constrained by multiple inter-corridor limit rules for unit commitment and a lower-level model based on the cluster affiliation criteria of the current operating mode. The two-tier optimization model is as follows:

$$\begin{aligned} &\min_{x} \ (1) \\ &\text{s.t.} \ (2),(4),(5) \end{aligned} \quad (11)$$

$$\Gamma_s(t) = \tilde{\partial}\Omega_j \pi_j(t), \forall j \in \mathbb{J} \quad (12)$$

$$\min_{\pi} \sum_{t=1}^{T}\sum_{j\in\mathbb{J}} \pi_j(t)\left[D_j^g(t)+D_j^w(t)+D_j^d(t)\right] \quad (13)$$

$$\text{s.t.} \sum_{j\in\mathbb{J}} \pi_j(t)=1, \pi_j(t)\in\{0,1\} \quad (14)$$

$$\begin{cases} D_j^g(t) = \sum_{g\in\mathbb{G}}\left(P_g(t)-P_{cg,j}\right)^2 \\ D_j^w(t) = \sum_{w\in\mathbb{W}}\left(P_w(t)-P_{cw,j}\right)^2 \\ D_j^d(t) = \sum_{d\in\mathbb{D}}\left(P_d(t)-P_{cd,j}\right)^2 \end{cases} \quad (15)$$

where $\pi_j(t)$ is a 0-1 integer variable that determines whether the operating mode belongs to the $j$-th sub-mode cluster, with 1 indicating membership and 0 indicating otherwise. The lower-level model aims to find the nearest class center to the current operating mode and assigns the index of this class center to the limit mapping rule set to activate the corresponding multiple inter-corridor limit rules. $P_{cg}$, $P_{cw}$, and $P_{cd}$ represent the synchronous machine, wind turbine unit output, and active load characteristics of the cluster center operating mode.

The above model is a mixed-integer two-level optimization model, with integer decision variables in the lower-level model. It can be transformed into a single-level model for a solution by relaxing the lower-level integer variables to continuous variables and utilizing KKT (Karush-Kuhn-Tucker) conditions. Introducing Lagrange multipliers, the Lagrangian function for the lower-level problem is given as follows:

$$\mathcal{L}(\pi,\alpha,\beta,\gamma) = \sum_{t=1}^{T}\left\{\begin{array}{l}\sum_{j\in\mathbb{J}}\pi_j(t)\left[D_j^g(t)+D_j^w(t)+D_j^d(t)\right]\\ +\alpha(t)(\sum_{j\in\mathbb{J}}\pi_j(t)-1)\\ +\sum_{j\in\mathbb{J}}\left[\beta_j(t)\pi_j(t)+\gamma_j(t)(\pi_j(t)-1)\right]\end{array}\right\} \quad (16)$$

The (16) can be transformed into constraints and embedded into the upper-level model for solving through KKT conditions[14].

## V. CASE STUDIES

The test case involves the IEEE 39-node system with two wind turbines added at nodes #17 and #21. The system structure and inter-corridor divisions are illustrated in Figure 2. We use a Python script to call the Gurobi solver for solving the inter-corridor limit switching optimization model. Safety checks encompass thermal stability, static voltage stability, and N-1 transient stability.

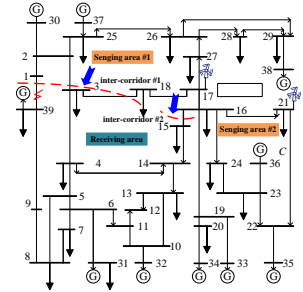

Figure 2. IEEE 39-bus system

### A. Effectiveness of Mutil−Inter-Corridor Limit Condition Mapping Rules

To validate the effectiveness of the proposed inter-corridor limit mapping rules in exploring inter-corridor potential, we set the levels of condition mapping to 2, 5, and 10, generating four different levels of inter-corridor limit mapping rules. We compare these rules to the most conservative limit calculation value of 500MW, as determined by traditional methods. The comparison results for the sending-end area 2 to the receiving-end inter-corridor are illustrated in Table 1.

TABLE I. MAXIMUM CONDITION MAPPING LIMIT FOR SENDING AREA 2 TO RECEIVING AREA UNDER SEVERAL BANDINGS

|  | Conservative model | Two-bandings | Five-bandings | Ten-bandings |
|---|---|---|---|---|
| Maximum limit/MW | 500 | 573.35 | 685.56 | 742.44 |

From Table I, it's clear that as the number of levels increases, the condition mapping set can accommodate larger maximum limits. In comparison to conservative limits, with 2, 5, and 10 levels, the maximum exploitable inter-corridor deliverability increases by 43%, 44.5%, and 48.4% respectively. This highlights that finer clustering of typical sub-operating modes allows for stronger potential for inter-corridor deliverability through condition mapping rules. In our test system, we've set the number of limit levels at 20 based on our testing experience. This value will be used unless otherwise specified in future cases.

### B. Validation of the Effectiveness of Mutil-Inter-Corridor Limit Switching Algorithm

To validate the potential advantages of the proposed multi-inter-corridor coupled limit rules in improving the integration of renewable energy generation and reducing dispatch costs, a comparative experiment was designed. It compared the inter-corridor export power under the proposed multi-inter-corridor coupled condition mapping limit constraints, independent inter-corridor condition mapping limit constraints, and traditional conservative limit constraints. The results, including the multi-inter-corridor limit boundaries and inter-corridor export power, are recorded in Figure 3.

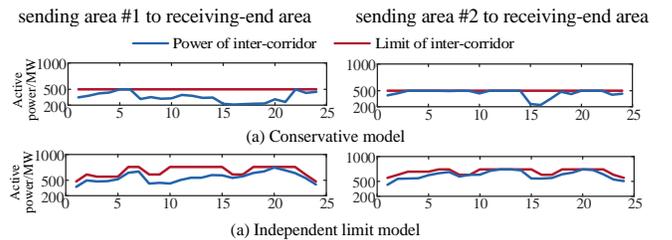

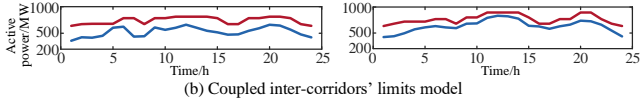
(b) Coupled inter-corridors' limits model

Figure 3. Multi-Inter-Corridors' Transfer Limits and Power

Figure 3 shows the visual results for two-inter-corridor limits and export power under three different methods. In all cases, inter-corridors' power flows are constrained by these limits. Compared to fixed conservative limit boundaries, both independent inter-corridor limits mapping and multi-inter-corridor coupled limit mapping models allow dynamic boundary switching based on the sequence of operating modes. They identify stable modes with higher inter-corridor export power, the main reason for unlocking inter-corridor potential. The proposed approach significantly improves inter-corridor deliverability. On average, it increases inter-corridor export power by 48.5% and transmission limits by 40% compared to traditional methods. This analysis demonstrates that the proposed approach effectively activates limit switching conditions, optimizing inter-corridor limits within the condition mapping set and successfully guiding unit commitment to operate more efficiently.

### C. Cross-Validation of Mutil-Inter-Corridor Limit Switching Decisions and Stability

To explore the potential correlations between the decision patterns of the proposed limit mapping constraint model and stability modes, the dispatch results of two models were selected at 6:00 AM: one without limit constraints and the other with the proposed limit constraints. These results were then verified for stability under a three-phase short-circuit fault condition.

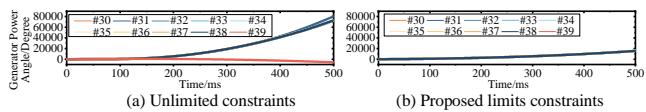

(a) Unlimited constraints    (b) Proposed limits constraints

Figure 4. Post-Fault Angular Trajectory

In Figure 4(a), the model without limit constraints results in transient angular instability, primarily affecting generator #39 while the other units remain synchronized. In the proposed limit constraints model, generator #39 is deactivated, but the remaining units stay synchronized, leading to a stable system, as shown in the trajectory in Figure 4(b). This demonstrates the flexibility of the proposed approach in switching between different unit commitments to find stable modes that enhance inter-corridor export power effectively.

### D. Enhanced Consumption of New Energy Resources and Improved Dispatch Cost

Table II presents the consumption rates of renewable energy during challenging periods(15:00) under four different methods. The proposed approach performs remarkably close to the ideal model without limits and significantly outperforms the other methods, confirming its effectiveness in harnessing inter-corridor export power potential.

TABLE II.  COMPARISONS ON RENEWABLE ENERGY CONSUMPTION RATE OF 4 RIVALS

| Consumption rate | Unlimits | Conservative limits | Independent limits | Proposed limits |
|---|---|---|---|---|
|  | 79.93% | 54.44% | 70.46% | 79.25% |

Table III compares the total scheduling costs of various comparison methods. Due to its ability to fully utilize inter-corridor export power potential, the proposed approach comes closest to the ideal model with no limit constraints in terms of unit commitment operating costs. This confirms its economic advantage.

TABLE III.  COMPARISONS ON UNIT COMMITMENT COST OF THE 4 RIVALS

|  | Unlimits | Conservative limits | Independent limits | Proposed limits |
|---|---|---|---|---|
| Total costs/$ | 2.876e10 | 3.052e10 | 2.885e10 | 2.879e10 |

### CONCLUSION

This paper proposes a targeted approach that includes a refined identification method for typical operating modes in renewable energy grids based on the ACC algorithm. Additionally, it presents an offline rule calculation method for coupling inter-corridors' limits based on condition mapping. Additionally, the paper proposes a solving paradigm employing the Lagrange multiplier method. Within the context of multi-inter-corridor limit adjustment, it suggests a method for establishing coordination rules for multi-inter-corridor limits. This forms the initial framework for a comprehensive system from limit setting calculation to optimization control. The case studies demonstrate that the proposed method effectively harnesses the potential of renewable energy export, balancing economic considerations, solution efficiency, and stability. It holds promise for theoretical research and practical engineering applications. Future work will involve further exploring the quantitative relationship between TTC and multi-inter-corridor limits, elucidating the computational principles behind inter-corridors' limits.

### REFERENCES

[1] M. Shaaban, W. Li, Z. Yan, Y. Ni, and F. F. Wu, "Calculation of total transfer capability incorporating the effect of reactive power," *Electric Power Systems Research*, vol. 64, no. 3, pp. 181–188, Mar. 2003.
[2] Y. Huang et al., "Linearized AC power flow model based interval total transfer capability evaluation with uncertain renewable energy integration," *International Journal of Electrical Power & Energy Systems*, vol. 154, p. 109440, Dec. 2023.
[3] L. Min and A. Abur, "Total Transfer Capability Computation for Multi-Area Power Systems," *IEEE Trans. Power Syst.*, vol. 21, no. 3, pp. 1141–1147, Aug. 2006.
[4] Gang Luo et al., "Fast calculation of available transfer capability in bulk interconnected grid," in *2012 IEEE Power and Energy Society General Meeting*, San Diego, CA: IEEE, Jul. 2012, pp. 1–8.
[5] W. Li, P. Wang, and Z. Guo, "Determination of optimal total transfer capability using a probabilistic approach," *IEEE Transactions on Power Systems*, vol. 21, no. 2, pp. 862–868, May 2006.
[6] G. Qiu et al., "Analytic Deep Learning-Based Surrogate Model for Operational Planning With Dynamic TTC Constraints," *IEEE Transactions on Power Systems*, vol. 36, no. 4, pp. 3507–3519, Jul. 2021.
[7] Y. Liu, J. Zhao, L. Xu, T. Liu, G. Qiu, and J. Liu, "Online TTC Estimation Using Nonparametric Analytics Considering Wind Power Integration," *IEEE Trans. Power Syst.*, vol. 34, no. 1, pp. 494–505, Jan. 2019.
[8] T. Levin and A. Botterud, "Capacity Adequacy and Revenue Sufficiency in Electricity Markets With Wind Power," *IEEE Trans. Power Syst.*, vol. 30, no. 3, pp. 1644–1653, May 2015.
[9] T. Ding, M. Qu, Z. Wang, B. Chen, C. Chen, and M. Shahidehpour, "Power System Resilience Enhancement in Typhoons Using a Three-Stage Day-Ahead Unit Commitment," *IEEE Trans. Smart Grid*, vol. 12, no. 3, pp. 2153–2164, May 2021.

[10] G. Chicco, O.-M. Ionel, and R. Porumb, "Electrical Load Pattern Grouping Based on Centroid Model With Ant Colony Clustering," *IEEE Transactions on Power Systems*, vol. 28, no. 2, pp. 1706–1715, May 2013.
[11] G. Chicco, R. Napoli, and F. Piglione, "Comparisons among clustering techniques for electricity customer classification," *IEEE Transactions on Power Systems*, vol. 21, no. 2, pp. 933–940, May 2006.
[12] H. Zhou, Z. Deng, Y. Xia, and M. Fu, "A new sampling method in particle filter based on Pearson correlation coefficient," *Neurocomputing*, vol. 216, pp. 208–215, Dec. 2016.
[13] J. Fürnkranz, D. Gamberger, and N. Lavrač, *Foundations of Rule Learning*. in Cognitive Technologies. Berlin, Heidelberg: Springer Berlin Heidelberg, 2012.
[14] Z. Zhao and L. Wu, "Impacts of High Penetration Wind Generation and Demand Response on LMPs in Day-Ahead Market," *IEEE Trans. Smart Grid*, vol. 5, no. 1, pp. 220–229, Jan. 2014.